\documentstyle[epsfig,12pt]{article}
\begin{document}
\newcommand{\3}{\ss}

\makebox[14cm][r]{PITHA 95/33}\par
\makebox[14cm][r]{November 1995}\par
\vspace{.7cm}
{\bf \centerline{\Large A prediction for three neutrino masses and mixings
and}}
\par
{\bf \centerline{\Large similarity between quark and lepton mixings}}
\vspace{1.cm}
\centerline{Saul Barshay and Patrick Heiliger {\footnote{Present address:
Union Investment, Westendstr. 1, D--60325 Frankfurt, Germany}}}
\par
\centerline{III. Physikalisches Institut (A), RWTH Aachen}
\par
\centerline{D--52056 Aachen, Germany}
\normalsize
\vspace{2.cm}

\begin{abstract}
If neutrinos have mass, we give reasons for a possible pattern of three
(squared) mass eigenvalues: $m_1^2 \simeq (2.8 - 5.8) \, \mbox{(eV)}^2 $,
$m_2^2 \simeq 0.01 \, \mbox{(eV)}^2 $, $m_3^2 \simeq (1.5 - 1) \times 10^{-4}
\mbox{(eV)}^2 $. The flavor states $\nu_{\mu} $ and $\nu_e $ are mixtures of
the eigen\-states with $m_2 $ and $m_3 $ with a significant mixing,
corresponding to an effective mixing angle of about 0.45. The $\nu_{\tau} $ is
nearly the state with $m_1 $; the other two effective mixing angles are about
an order of magnitude smaller than 0.45. There is a marked similarity to
mixing in the quark sector.
\end{abstract}

\newpage
The purpose of this note is to give reasons for the following speculative
prediction for three (Dirac) neutrino (squared) mass eigenvalues:
$m_1^2 \simeq 2.8 \, \mbox{(eV)}^2 $, $m_2^2 \simeq 0.01 \, \mbox{(eV)}^2 $,
and $m_3^2 \simeq 1.5 \times 10^{-4} \, \mbox{(eV)}^2 $. In addition one
predicts that $\nu_{\mu} $ and $\nu_e $ are mixtures of the eigenstates with
$m_2 $ (predominantly in $\nu_{\mu} $ via a $\cos \Theta $ factor) and $m_3 $
(predominantly in $\nu_e $) with a significant mixing, corresponding to an
effective mixing angle of about $\Theta = 0.45 $. The $\nu_{\tau} $ is nearly
the state with $m_1 $; the other two effective mixing angles are about an
order of magnitude smaller than 0.45. We show that this situation with
effective mixing angles in the lepton sector bears a strong resemblance to that
in the quark sector where $d $ and $s $ mix significantly via a Cabbibo angle
$\Theta_C \simeq 0.22 $, while the other two effective mixing angles bringing
in $b $ are much smaller. One can ask the question : \underline{why} is one
mixing angle significant and the others much smaller, and is there a relation
between the sizes ? We answer this question in both the lepton and quark
sectors by arguing that the significant effective mixing angle is the sum of
the quantities involving mass ratios, whereas the much smaller angles involves
the difference between such quantities of comparable size. One of our main
purposes here is thus to explicitly exhibit the potential numerical similarity
between \underline{all} mixings in the lepton and quark sectors.
 \par
\bigskip
We summarize first the empirical bases for our argument concerning a possible
set of connected values for three neutrino masses and their effective mixings
in three states of flavor. Central to the considerations in this paper is the
apparent difference between measured and predicted (muon/electron) flavor
composition in the atmospheric neutrino flux \cite{kamio}, \cite{imb},
\cite{soud}. This difference is not yet seen in some experiments \cite{frej},
 \cite{nusex}, but if it exists and is due to neutrino oscillations, then two
recent detailed analyses in particular \cite{mina}, \cite{fogli}, have shown
that there must exist a squared mass difference of about $10^{-2} \,
\mbox{(eV)}^2 $ and a significant mixing, corresponding to an effective mixing
angle $\Theta \ge 0.35 $. On the other hand, there exist definite indications
from analyses \cite{mina}, \cite{mina2} of different data, that two of the
three effective mixing angles are much smaller. We take as a second basis the
assumption that there exists a squared mass difference of about 3
$\mbox{(eV)}^2 $. This takes cosmological \cite{kolb} arguments for the
existence of some hot dark matter and assumes that at least one neutrino mass
contributes to this. \par
\bigskip
To define our notation, we give two approximate forms of the matrix which
gives the lepton flavor states in terms of the mass eigenstates. The full
matrix is that of the standard form \cite{rev} (eq. (28.3) in \cite{rev},
with $\delta_{13} = 0 $). We have $[ $F1$] $
\begin{eqnarray}
\left ( \begin{array}{ccc} \nu_e \\ \nu_{\mu} \\ \nu_{\tau} \end{array}
\right ) & \simeq &
\left ( \begin{array} {ccc} 0       & c_{13}        & s_{13} \\
                            -c_{23} & -s_{23}s_{13} & s_{23}c_{13} \\
                            s_{23}  & -c_{23}s_{13} & c_{23}c_{13}
        \end{array} \right )
\left ( \begin{array}{ccc} \nu_1(m_1) \\ \nu_2(m_2) \\ \nu_3(m_3) \end{array}
\right ) \nonumber \\
& \simeq & \left ( \begin{array} {ccc} c_{12}c_{13} & s_{12}c_{13} & s_{13} \\
                                      -c_{12}s_{13} &-s_{12}s_{13} & c_{13} \\
                                       s_{12}       &-c_{12}       & 0
                   \end{array} \right )
\left ( \begin{array}{ccc} \nu_1(m_1) \\ \nu_2(m_2) \\ \nu_3(m_3) \end{array}
\right )
\end{eqnarray}
The first form corresponds to setting $c_{12} = 0, s_{12} = 1 $; the second
corresponds to setting $c_{23} = 0, s_{23} = 1 $. Defining $\Theta_{12} =
\left ( {\pi \over 2} - \tilde{\Theta}_{12} \right ) $ and $\Theta_{23} =
\left ( {\pi \over 2} - \tilde{\Theta}_{23} \right ) $, this corresponds to
effective mixing angles $\tilde{\Theta}_{12} $ and $\tilde{\Theta}_{23} $
tending to zero, respectively. The reason for the interchange of the usual
roles of $\sin \Theta_{ij} $ ($\cos \Theta_{ij} $) close to zero (unity) is
simply because we have placed the largest--mass state $\nu_1(m_1) = \nu_1 $
uppermost in the column and the smallest--mass state $\nu_3 (m_3) = \nu_3 $
lowest. We do this in order to conform to the conventions utilized in the
recent analysis of data by Minakata \cite{mina} $[ $F2$] $; then our predicted
mass hierarchy corresponds to the allowed case $b $ following eq. (15) in
\cite{mina}. We also define $\Theta_{13} = \left ( {\pi \over 2} -
\tilde{\Theta}_{13} \right )$. It is our result below, $\tilde{\Theta}_{13}
\simeq 0.45 $, which gives rise to a signifcant mixing of $\nu(m_2) = \nu_2 $
and $\nu (m_3) = \nu_3 $ in the states $ \nu_{\mu} $ and $\nu_e $. The
theoretical speculation which we utilize is that the significant effective
mixing angle $\tilde{\Theta}_{13} $, and the much smaller angles
$\tilde{\Theta}_{12}, \tilde{\Theta}_{23} $ are given by the following
equations, respectively, in terms of mass ratios.
$$
\;\;\;\;\;\;\;\;\;\;\;\;\;\;
\tilde{\Theta}_{13}  \simeq  \left \{ \left ({m_e \over m_{\mu}}
\right )^{1/2} + \left ( {m_3 \over m_2} \right )^{1/2} \right \}
 \simeq  \{0.07 + 0.38 \} \simeq 0.45  \;\;\;\;\;\;\;\;\;\;\;\;\;\;\;\;\;\;
\;\;\;\;\;\;\; (2a)
$$
\begin{eqnarray}
\;\;\;\;\;\;\;\;\;\;\;\;\;\;
\tilde{\Theta}_{12} \sim \tilde{\Theta}_{23} \simeq  \left \{ \left (
{m_{\mu} \over m_{\tau}} \right )^{1/2} - \left ({m_2 \over m_1} \right )^{1/2}
\right \}   \approx \{0.245 - 0.245 \} \approx 0 \;\;\;\;\;\;\;\;\;\;\;\;\;\;
\; (2b) \nonumber
\end{eqnarray}
Starting with the indication from the atmospheric neutrino anomaly, $m_2 \simeq
\sqrt{0.01} $ eV in eq. (2b), $m_1 \sim \sqrt{2.8} $ eV is estimated by taking
$\tilde{\Theta}_{12}, \tilde{\Theta}_{23} $ as tending to zero. This is. of
course, an idealization. A small value for these angles $\leq 0.04 $ arises for
$m_1 \leq \sqrt{5.8} $ eV. Then, $m_3 \simeq \sqrt{1.5 \cdot 10^{-4}} $ eV in
eq. (2a) is estimated by taking the ratio of ratios ${(m_2/m_1)^{1/2} \over
(m_3/m_2)^{1/2}} \sim 0.65 $, for \underline{orientation}. This is the same
number that occurs in the quark sector (note the ratio $(0.146/0.224) $ in eqs.
(4a, b) below). With this number, for $m_1 \leq \sqrt{5.8} $ eV, we have
$m_3 \geq \sqrt{10^{-4}} $ eV. Its is noteworthy that if the minus sign
between the two terms in eq. (2b) were to be changed to a plus sign, the sum
would be $\sim 0.49 $, nearly the same number as in eq. (2a). With the minus
sign, a near cancellation of two comparable terms occurs. \par
\bigskip
For comparison, we apply these considerations to the quark sector, using
\cite{rev} $[ $F3$] $ $m_u \simeq 2 $ MeV, $m_d \simeq 5 $ MeV, $m_c \simeq
1.55 $ GeV, $m_s \simeq 100 $ MeV, $m_t \simeq 180 $ GeV, $m_b \simeq 4.7
$ GeV. Using the standard mixing matrix \cite{rev}, with $s_{13} \rightarrow
0 $, the flavor states (primed) are
\begin{eqnarray}
\;\;\;\;\;\;\;\;\;\;\;\;\;\;\;\;\;\;\;
\left ( \begin{array} {ccc} d' \\ s' \\ b' \end{array} \right ) \simeq
\left ( \begin{array} {ccc} c_{12}       & s_{12}        &    0   \\
                           -c_{23}s_{12} & c_{23}c_{12}  & s_{23} \\
                            s_{23}s_{12} & -s_{23}c_{12} & c_{23}
        \end{array} \right )
\left ( \begin{array} {ccc} d \\ s \\ b \end{array} \right )
\;\;\;\;\;\;\;\;\;\;\;\;\;\;\;\;\;\;\;\;\;\;\;\;\;\;\;\;\;\;\;\;\;\;\;\;
(3) \nonumber
\end{eqnarray}
The equations for effective mixing angles analogous to eqs. (2a, b) are
$$
\;\;\;\;\;\;\;\;\;\;\;\;\;\;
\Theta_{12} = \left \{ \left ( {m_u \over m_c} \right )^{1/2} + \left (
{m_d \over m_s} \right )^{1/2} \right \} \simeq \{0.036 + 0.224 \} \simeq
0.26 \;\;\;\;\;\;\;\;\;\;\;\;\;\;\;\;\;\;\;\;\;\;\; (4a)
$$
$$
\;\;\;\;\;\;\;\;\;\;\;\;\;\;
\Theta_{23} = \left \{ - \left( {m_c \over m_t} \right )^{1/2} + \left (
{m_s \over m_b} \right )^{1/2} \right \} \simeq \{-0.093 + 0.146 \}
\simeq 0.05 \;\;\;\;\;\;\;\;\;\;\;\;\;\;\;\;\;\;  (4b)
$$
The closeness of the effective mixing angle in eq. (4a) (specifically,
$(m_u/m_s)^{1/2} \simeq 0.224 $) to the empirical Cabibbo angle is, of course,
known \cite{weinb}. However, the near cancellation of two comparable terms in
eq. (4b) results in a significantly smaller angle. This is close to the
empirical value (note eq. (28.2) in \cite{rev}); it is closer than the
quantity $(m_s/m_b)^{1/2} $ alone. If the top mass were nearer to the
intermediate vector--boson masses, $\approx $ 90 GeV, eq. (2b) would give a
mixing angle which approaches zero. Again it is noteworthy that if the minus
sign in eq. (4b) were to be changed to a plus sign, the sum would be 0.24,
nearly the same number as in eq. (4a). Moreover, there is a marked similarity
between the overall \underline{pattern} of the four numbers, square roots of
mass ratios, involved in the addition and subtraction in eqs. (4a, b) and the
hypothetical pattern of the four numbers involved in the addition and
subtraction in eqs. (2a, b). This is reflected in the scale factor of only
$\sim 1.7 $ difference between the sum in eq. (2a) from that in eq. (4a), and
in the similar smallness of the differences in eq. (2b) and eq. (4b). In this
sense, the mixing of mass eigenstates in the lepton and quark sector are not
as different as has often seemed. \par
\smallskip
These numerical similarities are interesting to observe, even in the absence
of a detailed theoretical model. The origin of relations like those in eqs.
(2a,b) and eqs. (4a,b) can be in dynamical models in which the lower masses
are generated in second order of a Higgs--type or a $\sigma $--model type
\cite{saul} mixing interaction to the mass above, that is $m_2 \sim g_{21}^2
m_1, m_3 \sim g_{32}^2 m_2 $. The sign of the couplings $g_{21} $ and $g_{32}
$ are physically relevant in determining the addition or subtraction for
calculating the effective mixing angles. Of course, the ``starting'' mass
value, i.e.the extremely small value of $m_{\nu_{\tau}} \sim 1.7 $ eV relative
to $m_{\tau} \simeq 1.8 $ GeV is not explained. (The mass ratio $\sim 10^9 $
is maintained for $m_{\nu_{\mu}} \sim 0.1 $ eV, $m_{\mu} \simeq 100 $ MeV.)
Explanations of this (i.e. a ``see--saw'' type model) usually introduce a
new mass scale of about $10^9 $ GeV (see eq. 11.19 in \cite{har}). However,
one can note that in the neutrino mass pattern postulated here, the ratios
$(m_{\mu} / (m_2 \cdot 10^6)) $, $(m_{\tau} / (m_1 \cdot 10^6)) \simeq 10^3 $
have increased by a factor of $\sim 25 $ over the first ratio $(m_e / (m_3
\cdot 10^6)) \simeq 40 $. In the quark sector, the ratios $(m_c/m_s ) \simeq
15.5 $ and $(m_t/m_b) \simeq 38.5 $ represent increase factors which are
comparable to 25, over the ratio $(m_u/m_d) $ taken as $\sim 1 $. The relevant
factor of $10^{-9} $ might, for example, be associated with a small admixture
$r^2 \geq 10^{-9} $ of right--handed coupling giving $m_3 \sim r^2 m_{\tau} $
as a weak radiative effect. \par
\bigskip
We have not explained the magnitude of the apparent deficiency of solar
neu\-trinos with our hypothetical neutrino mass and mixing pattern. However,
vacuum oscillations for the relatively large relevant $\Delta m^2 \sim
10^{-2} (\mbox{eV})^2 $ do lead to some deficiency, at the $30 \, \% $ level.
Other possibilities (i.e. MSW mixing) are summarized in the vertical lines in
figs. (4a, b) and figs. (5a, b) of \cite{shi}, which also discusses the effect
of a cooler sun. \par
\bigskip
We note that some speculations \cite{mina}, \cite{cald} have
concentrated upon a hypothesis for the existence of two nearly degenerate
neutrino mass eigenstates with mass $\sim (6.5 - 3.5) $ eV and a degeneracy at
the level of $10^{-3} $ eV. These must mix significantly $[ $F4$] $ to form
$\nu_{\mu} $ and $\nu_{\tau} $ $[ $F5$] $. Then the atmospheric neutrino
anomaly involves $\nu_{\mu} \to \nu_{\tau} $ rather than $\nu_{\mu} \to \nu_e $
as in the mass pattern given in this paper. There is no explanation of the
apparent solar neutrino deficiency with only the three known neutrinos. \par
\bigskip
There are also recent speculations \cite{val}, \cite{fri} that the lightest
neutrino is also nearly degenerate with the two: with a degree of degeneracy
of about \cite{val} $10^{-6} $ eV, or with a remarkable degree of degeneracy of
about \cite{fri} $10^{-11} $ eV. Such degrees of degeneracy are postulated
in order to interpret all of the experiments on the solar neutrino defficiency:
in the former case \cite{val} via matter--enhanced oscillations with a very
small effective mixing angle, and in the latter case \cite{fri} via vacuum
oscillations with a large mixing angle. \par
\bigskip
In conclusion, we note the implications of the present hypothesis for certain
experiments which are continuing, and put them in the context of recent
analyses. There is, in principle a mixing effect in the LSND experiment for
$\overline{\nu}_{\mu} \to \overline{\nu}_e $ \cite{lsnd}. This occurs
effectively via a ``virtual $\overline{\nu}_{\tau} $''; this effect was
analyzed by Minakata \cite{mina} and has also been discussed by others
\cite{babu}. However, with $\tilde{\Theta}_{12}, \tilde{\Theta}_{23} < 0.07 $,
the probability is more than an order of magnitude below the possible effect
in the first paper of \cite{lsnd}. The $(\Delta m^2 - sin^2 {2
\tilde{\Theta}_{13}} ) $ values discussed here lie just inside $[ $F 6$] $ the
region allowed by reactor experiments \cite{achk}, \cite{vid}; this is best
seen by observing the allowed region relevant to the possibility of
atmospheric $\nu_{\mu} - \nu_e $ mixing in the fig. 3 of the recent analysis
\cite{bil} of atmospheric neutrino oscillations $[ $F 7$] $. Thus, an effect
should emerge in the on--going reactor experiments, and in the up--coming
super--Kamiokande experiment \cite{bil}. \par
\bigskip
\bigskip
\noindent
{\bf Acknowledgement} \par
One author (S. B.) wishes to thank Prof. J.F.W. Valle for most helpful
conversations in Valencia.

\newpage
\large
\noindent
{\bf Footnotes}
\normalsize
\begin{itemize}
\item[[F 1]] The flavor states are thus transparently given by
             \begin{eqnarray} |\nu_e > & = & c_{13} \, |\nu_2 > + \, s_{13}
             \, |\nu_3 > \nonumber \\
             |\nu_{\mu} > & = & -c_{23} \, |\nu_1 > + \, s_{23} \,
             |\nu_{\mu} ' > \nonumber \\
             |\nu_{\tau} > & = & s_{23} \, |\nu_1 > + \, c_{23} \,
             |\nu_{\mu} ' > \nonumber \end{eqnarray}
             with $|\nu_{\mu} ' > = -s_{13} \, |\nu_2 > + c_{13} |\nu_3 > $.
\item[[F 2]] Minakata emphasizes that the analysis of data in \cite{mina}
             does not depend upon the relative magnitude of $m_2 $ and $m_3 $;
             it depends upon the difference in squared mass.
\item[[F 3]] Consistently, we use the lower values for $m_u, m_d $ and $m_s $,
             with $(m_u/m_d) $ and $(m_s/m_d) $ in the middle of their allowed
             range (see pages 1437, 1438 in \cite{rev}).
\item[[F 4]] One may note that a relation like eq. (2b) can accomodate a
             maximal effective mixing angle for $(m_2/m_1) \sim 1 $, namely
             $$\Theta \simeq \left \{ - \left ({m_{\mu} \over m_{\tau}}
             \right )^{1/2} + \left ({m_2 \over m_1} \right )^{1/2} \right \}
             \simeq \{-0.25 + 1\} \simeq 0.75 $$.
\item[[F 5]] The hypothetical mass degeneracy avoids the stringent limits set
             for $\nu_{\mu} \to \nu_{\tau} $ mixing when $\Delta m^2 $ is
             greater than about $3 (\mbox{eV})^2 $ by the experiments in
             \cite{e531}. (It has been stated that such mass degeneracy might
             be ``natural'' for Majorana neutrinos \cite{wolf}.)
\item[[F 6]] S. B. thanks Jos\'e Valle for emphasizing this point to him.
\item[[F 7]] Note that the index 3 in \cite{bil} is 2 in the present context.
             In \cite{bil}, the heaviest mass (denoted by 3) was, in effect,
             assumed to be no greater than a few tenths of an electron-volt.
             Thus making no significant contribution to dark matter.
\end{itemize}

\end{document}